\documentclass[conference]{IEEEtran}
\IEEEoverridecommandlockouts
\usepackage{cite}
\usepackage{amsmath,amssymb,amsfonts}
\usepackage{algorithmic}
\usepackage{authblk}
\usepackage{float}  % 放在导言区
\usepackage{graphicx}
\usepackage{textcomp}
\usepackage{xcolor}
\usepackage{subfigure}
\usepackage{multirow} 
\usepackage{booktabs}
\usepackage{adjustbox}
\def\BibTeX{{\rm B\kern-.05em{\sc i\kern-.025em b}\kern-.08em
    T\kern-.1667em\lower.7ex\hbox{E}\kern-.125emX}}
\usepackage{algorithmic}
\usepackage{algorithm}

\usepackage{hyperref}
\usepackage[capitalize]{cleveref}

\usepackage{xspace}
\makeatletter
\DeclareRobustCommand\onedot{\futurelet\@let@token\@onedot}
\def\@onedot{\ifx\@let@token.\else.\null\fi\xspace}

\def\eg{\emph{e.g}\onedot}

\def\etal{\emph{et al}\onedot}
\makeatother

\title{
TDADL-IE: A Deep Learning-Driven Cryptographic Architecture for Medical Image Security
}
\author[1]{Junhua Zhou}
\author[1]{Quanjun Li}
\author[1]{Weixuan Li}
\author[2,3,4]{Guang Yu}
\author[4]{YiHua Shao}
\author[2]{Yihang Dong}
\author[5]{\authorcr Mengqian Wang}
\author[2*]{Zimeng Li\thanks{* Corresponding authors: li\_zimeng@szpu.edu.cn, xuhangc@hzu.edu.cn}}
\author[2]{Changwei Gong}
\author[2*]{Xuhang Chen\thanks{This work was supported in part by the National Natural Science Foundation of China (Grant No. 62501412), in part by Shenzhen Medical Research Fund (Grant No. A2503006), in part by Shenzhen Polytechnic University Research Fund (Grant No. 6025310023K) and in part by Guangdong Basic and Applied Basic Research Foundation (Grant No. 2024A1515140010).}}

\affil[1]{Guangdong University of Technology}
\affil[2]{School of Electronic and Communication Engineering, Shenzhen Polytechnic University}
\affil[3]{School of Computer Science and Engineering, Huizhou University}
\affil[4]{University of Science and Technology Beijing}
\affil[5]{Institute of Neuroscience, Chinese Academy of Sciences}

\begin{document}
\maketitle

\begin{abstract}
The rise of digital medical imaging, like MRI and CT, demands strong encryption to protect patient data in telemedicine and cloud storage. Chaotic systems are popular for image encryption due to their sensitivity and unique characteristics, but existing methods often lack sufficient security. This paper presents the Three-dimensional Diffusion Algorithm and Deep Learning Image Encryption system (TDADL-IE), built on three key elements. First, we propose an enhanced chaotic generator using an LSTM network with a 1D-Sine Quadratic Chaotic Map (1D-SQCM) for better pseudorandom sequence generation. Next, a new three-dimensional diffusion algorithm (TDA) is applied to encrypt permuted images. TDADL-IE is versatile for images of any size. Experiments confirm its effectiveness against various security threats. The code is available at \href{https://github.com/QuincyQAQ/TDADL-IE}{https://github.com/QuincyQAQ/TDADL-IE}.
\end{abstract}


\section{INTRODUCTION}
\label{sec1}
The rise of medical imaging and telemedicine has increased the need for secure digital image transmission and storage. Medical images like MRI, CT, and ultrasound contain sensitive patient data crucial for diagnosis and treatment. Without strong encryption, these images are vulnerable to unauthorized access and breaches, especially over unsecured networks or in cloud storage. Protecting this data is essential for patient privacy and required by regulations like HIPAA. Traditional text-oriented encryption methods are ineffective for medical images due to their high inter-pixel correlation, large data size, and complex formats. Chaotic systems offer a promising solution for developing robust and efficient image encryption techniques \cite{chen2019medical}.

Chaotic systems, known for their deterministic randomness, ergodicity, and sensitivity to initial conditions, are ideal for cryptographic applications like generating pseudo-random sequences for image encryption \cite{xiaofu1999general}. Although one-dimensional (1D) chaotic maps are simple and easy to implement \cite{huang2019symmetric}, their predictability and limited chaos can create security risks \cite{xiaofu1999general}. Thus, research has turned to more complex chaos-based encryption systems. Recently, deep learning has emerged as a new approach, with efforts like Google Brain's self-encryption model using adjoint networks \cite{abadi2016learning}, Long Short-Term Memory (LSTM) networks for assessing chaotic sequence randomness \cite{he2021new}, and hybrid neural-chaotic encryption methods \cite{wei2020general, zhao2021new}. Despite these progressions, deep learning in color image encryption, with its challenges of inter-channel dependencies and high dimensionality, remains underexplored.

Permutation-only encryption schemes, common in image encryption, are vulnerable to various attacks due to discernible statistical properties in ciphertext \cite{jolfaei2015security}. Diffusion processes modify pixel values to enhance encryption by obscuring local dependencies. Many existing diffusion schemes lack sensitivity to plaintext and key variations as they operate independently of image content. Recent research has focused on plaintext-aware diffusion \cite{murillo2015rgb,parvin2016new,wu2017novel,li2017plaintext}. Murillo et al. used logical maps based on plaintext features \cite{murillo2015rgb}, and Parvin et al. combined XOR operations with chaotic functions \cite{parvin2016new}. However, these methods rely on initial plaintext for key generation, limiting their use in real-time scenarios like live video encryption. Their weak nonlinearity and limited key spaces make them susceptible to chosen-plaintext attacks \cite{fan2018cryptanalysis,norouzi2016breaking}. To address these issues, we propose the Joint Nonlinear Three-dimensional Diffusion Mechanism (TDA), which tightly integrates diffusion with the image content, enhancing sensitivity and avalanche effects.

Our primary contributions are as follows:
\begin{enumerate}

\item Existing one-dimensional chaotic maps frequently suffer from constrained chaotic ranges and suboptimal sensitivity to initial conditions, thereby limiting their efficacy in generating secure cryptographic keys. Therefore, a novel 1D-chaotic map 1D-Sine Quadratic Chaotic Map (1D-SQCM) is proposed, which can generate sequences with enhanced randomness and wider chaotic ranges.

\item By synergistically integrating the 1D-Sine Quadratic Chaotic Map (1D-SQCM) with the dynamic modeling capabilities of Long Short-Term Memory (LSTM) networks, we engineer a novel chaotic sequence generator. This generator exhibits an expanded chaotic regime, superior entropy, and refined temporal correlations, culminating in substantially enhanced randomness and cryptographic robustness.

\item Addressing the distinct requirements of diffusing non-square and color images, where inter-channel correlations and varying dimensions pose significant challenges, we present the three-dimensional diffusion algorithm (TDA). This mechanism amalgamates forward and backward diffusion processes with inter-channel pixel interactions. Such a design ensures that minute alterations in the plaintext cascade comprehensively throughout the ciphertext, thereby significantly amplifying both key sensitivity and plaintext sensitivity.
\end{enumerate}

\section{METHODOLOGY}
\label{Sec2}
\begin{figure}[ht]
    \centering
    \includegraphics[width=1\linewidth]{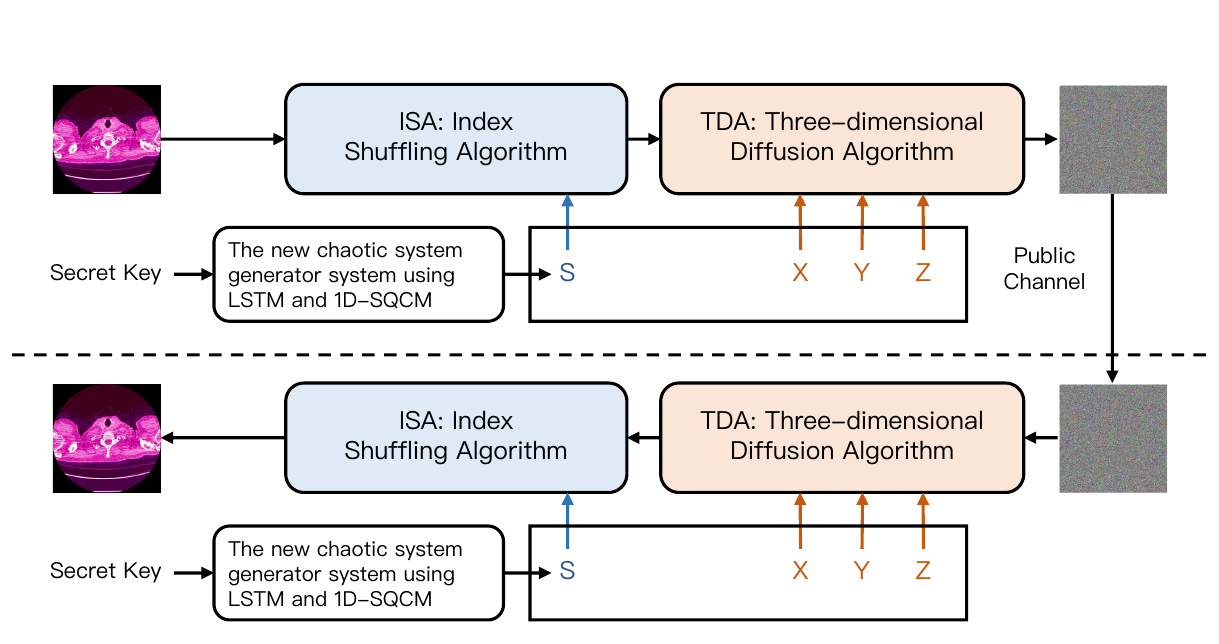}
    \caption{The process of the encryption algorithm.}
    \label{process}
\end{figure}

TDADL-IE is a new encryption algorithm for multi-medical images, using a 1D Sine-Cosine-Exponential Map (1D-SQCM) and neural network enhancement. Its flowchart is shown in \Cref{process}. The algorithm has two main modules: (1) chaotic sequence generation with 1D-SQCM and BLSTM network, and (2) a Three-dimensional Diffusion Algorithm (TDA). TDADL-IE can encrypt multiple images simultaneously. The secret keys for encryption are $Key_{x1}$, $Key_{y1}$, $Key_{a}$, and $Key_{N_0}$. The modules are detailed in the following subsections.

\subsection{Chaotic Sequence Generation using 1D-SQCM and BLSTM}
\label{section2}

\begin{figure}[ht]
	\centering
	\includegraphics[width=\columnwidth]{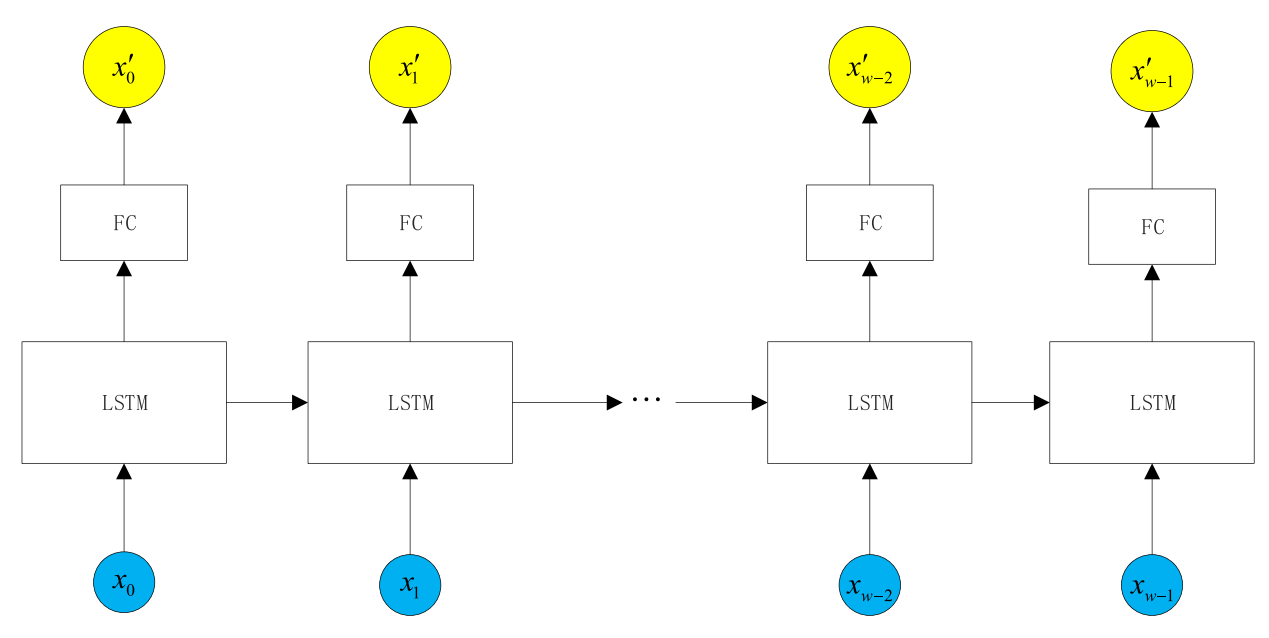}
	\caption{LSTM training model diagram. a Training diagram of x}
	\label{LSTM training model diagram}
\end{figure}

Bidirectional Long Short-Term Memory (BLSTM) networks enhance standard LSTMs \cite{hochreiter1997long} with a design involving input, forget, and output gates, and a memory cell (LSTM structure shown in \cref{LSTM training model diagram}). The forget gate $f_t$ regulates retaining information from previous states $h_{t-1}$ and its output is limited to $[0,1]$. In generating chaotic-like sequences, the forget gate's selective discarding of historical information mirrors the fading memory of chaotic systems, increasing sensitivity to initial conditions and encouraging divergent phase-space trajectories. Simultaneously, the input gate governs new information input, introducing nonlinear changes that influence system evolution and enhance complex dynamics. The memory cell integrates these inputs, updating its state to capture long-term dependencies and produce intricate, high-dimensional sequences. Altogether, these elements form a robust nonlinear mechanism for modeling and creating complex temporal sequences with chaotic traits.

Leveraging these properties, we introduce a hybrid chaotic sequence generator that synergistically combines the 1D Sine-Cosine-Exponential Map (1D-SQCM) with the temporal modeling capabilities of a BLSTM network. The 1D-SQCM, selected for its inherent characteristics of high entropy and pronounced sensitivity to initial conditions, is defined as:
\begin{equation}
	\label{TICMIC}
	x_{n+1}=\sin(\frac{a^2 }{\sin(x_{n})}),
\end{equation}
where $x_n$ denotes the state variable at iteration $n$, and $a$ is a system control parameter. By subsequently processing the output of the 1D-SQCM through the BLSTM network, we aim to generate sequences that are not only temporally correlated and statistically rich but also exhibit significant phase space divergence. This integrated methodology preserves the fundamental randomness of the chaotic map while empowering the BLSTM to adaptively learn and amplify the complexity and unpredictability of the generated sequences.

\textbf{Step 1: BLSTM Model Training.}
The BLSTM model undergoes an initial training phase to learn the complex dynamics of chaotic sequences. This process is initiated as follows:
\begin{enumerate}
\item \textit{Training Data Generation}: The system parameter $Key_a$ for the 1D-SQCM (\cref{TICMIC}) and an initial state $Key_{x0}$ (distinct from the encryption keys used for image processing) are specified. The 1D-SQCM is iterated to produce a raw chaotic sequence. To mitigate transient effects and ensure the sequence operates within the chaotic attractor, the initial $N_{\text{discard\_train}} = 6,000$ iterates are discarded. The subsequent $N_{\text{train}} = 24,000$ points are collected to form the training sequence, denoted as ${X_{\text{train}}}$.
\item \textit{BLSTM Configuration and Training}: The architecture of the BLSTM network is defined, including the number of hidden layers and units. Key hyperparameters, such as the learning rate and the length of input subsequences for training, are optimized. The BLSTM model is then trained using ${X_{\text{train}}}$ to learn its underlying temporal patterns and statistical properties.
\end{enumerate}
Upon completion of this training phase, the BLSTM model is equipped to generate enhanced chaotic sequences for the encryption process.

\textbf{Step 2: Hybrid Chaotic Sequence Generation for Encryption.}
For the encryption process, the secret keys $Key_{x1}$, $Key_{y1}$, $Key_{a}$, and $Key_{N_0}$ are employed. The generation proceeds as follows:
\begin{enumerate}
\item \textit{1D-SQCM Sequence Generation}: A primary chaotic sequence, denoted ${X_{\text{SQCM}}}$, of length $2mn$ is generated using the 1D-SQCM (\cref{TICMIC}). The map is initialized using $Key_{x1}$ as the initial state $x_0$, and $Key_a$ serves as the control parameter. The keys $Key_{y1}$ and $Key_{N_0}$ may influence aspects such as the number of initial iterations to discard prior to collecting sequence values for encryption.
\item \textit{BLSTM Sequence Enhancement}: The trained BLSTM model is then utilized to produce a secondary chaotic sequence, ${X_{\text{BLSTM}}}$, also of length $2mn$. This sequence can be generated by feeding an initial seed (derived from the keys or from a segment of ${X_{\text{SQCM}}}$) to the BLSTM and allowing it to predict subsequent values.
\item \textit{Sequence Concatenation}: The two sequences, ${X_{\text{SQCM}}}$ and ${X_{\text{BLSTM}}}$, are horizontally concatenated to form a hybrid chaotic sequence ${S_{\text{hybrid}}}$ of total length $4mn$. This hybrid sequence benefits from both the fundamental chaotic properties of the 1D-SQCM and the learned complexity introduced by the BLSTM.
\end{enumerate}
The length $mn$ typically corresponds to the number of pixels in a single channel of an image of size $m \times n$. The total length $4mn$ prepares for processing data equivalent to an $m \times n$ image with four components per pixel (\eg, RGBA) or four distinct $m \times n$ sequences for cryptographic operations.

\textbf{Step 3: Chaotic Sequence Processing and Derivation.}
The hybrid chaotic sequence ${S_{\text{hybrid}}}$ of length $4mn$, generated as described in Step 2, is subsequently processed to derive tailored sequences for the various stages of the encryption algorithm. This processing is performed as follows:
\begin{enumerate}
\item \textit{Sequence Partitioning}: The sequence ${S_{\text{hybrid}}}$ is partitioned into four sub-sequences, ${s}$, ${x}$, ${y}$, and ${z}$, each of length $L=mn$. (Here, $L=mn$ represents the number of elements in each sub-sequence, corresponding to the pixel count of a single-channel $m \times n$ image).
\item \textit{Derivation of Diffusion Sequences}: From the sub-sequences ${x}$, ${y}$, and ${z}$, three new sequences, $X$, $Y$, and $Z$, also of length $L=mn$, are computed using the transformations defined in \cref{pre-random}.
\begin{equation}
\begin{aligned}
	\label{pre-random}
	&\left\{
	\begin{aligned}
		X(i) &= \text{mod}(\lfloor x(i) \cdot y(L\!-\!i\!+\!1) \cdot 10^9 \rfloor, 256) + 1, \\
		Y(i) &= \text{mod}(\lfloor x(L\!-\!i\!+\!1) \cdot y(i) \cdot 10^9 \rfloor, 256) + 1, \\
		Z(i) &= \text{mod}(\lfloor z(i) \cdot 10^9 \rfloor, 256) + 1,
	\end{aligned}
	\right.
\end{aligned}
\end{equation}
for $i = 1, 2, \dots, L$. 

\item \textit{Matrix Reshaping}: The resulting one-dimensional sequences $X$, $Y$, and $Z$ are reshaped into $m \times n$ matrices, denoted $X_{\text{2D}}$, $Y_{\text{2D}}$, and $Z_{\text{2D}}$, respectively. These matrices are subsequently employed in the TDA module. The sequence ${s}$ is reserved for the Index Shuffling Algorithm.
\end{enumerate}
This multi-stage derivation ensures that distinct, complex pseudo-random sequences are available for different cryptographic operations, enhancing the overall security.

\subsection{Index Shuffling Algorithm (ISA)}
\label{subsec:ISA}
The Index Shuffling Algorithm (ISA) implements a pixel-level permutation to obfuscate spatial correlations within the input image. This process is applied to each $m \times n$ image or image channel independently, using the chaotic sequence ${s}$ (of length $mn$) derived as described in \cref{section2} (Step 3). The ISA proceeds as follows:
\begin{enumerate}
\item \textit{Vectorization}: The $m \times n$ pixel data of an image (or an individual channel thereof) is flattened into a one-dimensional vector $P$ of length $mn$.
\item \textit{Permutation Index Generation}: The chaotic sequence ${s}$ is utilized to derive a set of permutation indices. A common method involves obtaining the indices that would sort the elements of ${s}$. Let these indices be $Idx = [\text{idx}_1, \text{idx}2, \dots, \text{idx}{mn}]$.
\item \textit{Pixel Scrambling}: The pixels in vector $P$ are reordered according to these indices. Specifically, the $j$-th pixel in the permuted vector $P'$ is $P(\text{idx}_j)$. This operation effectively scrambles the original pixel arrangement.
\item \textit{Reshaping}: The permuted vector $P'$ is reshaped back into its original $m \times n$ two-dimensional structure, yielding the shuffled image (or channel).
\end{enumerate}
This permutation step significantly enhances the algorithm's resistance to statistical attacks by decorrelating adjacent pixels.

\subsection{Three-dimensional Diffusion Algorithm (TDA)}
\label{section3}

\begin{algorithm}[ht]
    \caption{Three-dimensional Diffusion Algorithm}
    \label{Alg1}
    \begin{algorithmic}[1]
        \REQUIRE The scrambled image $A$ with size $M\times N$ and the chaotic matrices $X_{2D}$, $Y_{2D}$, and $Z_{2D}$
        
        \STATE $C(1,1,1) \gets mod(A(1,1,1) + X_2D(1,1), 256)$
        \STATE $C(1,1,2) \gets mod(A(1,1,2) + Y_2D(1,1), 256)$
        \STATE $C(1,1,3) \gets mod(A(1,1,3) + Z_2D(1,1), 256)$
        
        \FOR{$j=2$ to $N$}
            \STATE $C(1,j,1) \gets mod(A(1,j,1) + X_2D(1,j) + C(1,j-1,1) + C(1,j-1,3), 256)$
            \STATE $C(1,j,2) \gets mod(A(1,j,2) + Y_2D(1,j) + C(1,j-1,2) + C(1,j-1,1), 256)$
            \STATE $C(1,j,3) \gets mod(A(1,j,3) + Z_2D(1,j) + C(1,j-1,3) + C(1,j-1,2), 256)$
        \ENDFOR
        
        \FOR{$i=2$ to $M$}
            \STATE $C(i,1,1) \gets mod(A(i,1,1) + X_2D(i,1) + C(i-1,1,1) + C(i-1,1,3), 256)$
            \STATE $C(i,1,2) \gets mod(A(i,1,2) + Y_2D(i,1) + C(i-1,1,2) + C(i-1,1,1), 256)$
            \STATE $C(i,1,3) \gets mod(A(i,1,3) + Z_2D(i,1) + C(i-1,1,3) + C(i-1,1,2), 256)$
        \ENDFOR
        
        \REPEAT
            \FOR{$i=2$ to $M$}
                \FOR{$j=2$ to $N$}
                    \STATE $C(i,j,1) \gets mod(A(i,j,1) + X_2D(i,j) + C(i,j-1,3) + C(i-1,j,3), 256)$
                    \STATE $C(i,j,2) \gets mod(A(i,j,2) + Y_2D(i,j) + C(i,j-1,1) + C(i-1,j,1), 256)$
                    \STATE $C(i,j,3) \gets mod(A(i,j,3) + Z_2D(i,j) + C(i,j-1,2) + C(i-1,j,2), 256)$
                \ENDFOR
            \ENDFOR
        \UNTIL{The image encryption process is complete.}
        
        \STATE \textbf{Output:} The encrypted image $C$
    \end{algorithmic}
\end{algorithm}

The Three-dimensional Diffusion Algorithm (TDA) is designed to spread pixel information across spatial dimensions and color channels in images $m \times n$. It uses forward and backward diffusion passes to ensure changes to a single plaintext pixel impact many ciphertext pixels, ensuring diffusion. The forward pass processes pixels in raster order \eg from top-left $(1,1)$ to bottom-right $(m,n)$, while the backward pass goes in the reverse direction \eg from $(m,n)$ to $(1,1)$.

\begin{figure}
	\centering
	\includegraphics[width=0.82\columnwidth]{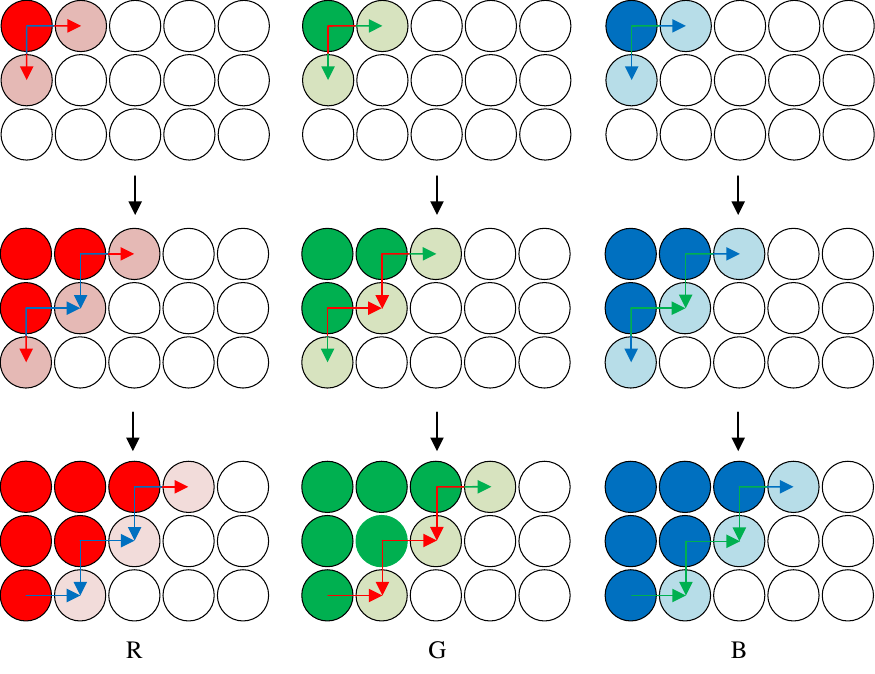}
	\caption{The Illustration of Three-dimensional Diffusion Algorithm (TDA)}
	\label{3d_fission}
\end{figure}

TDA's cross-channel pixel interactions boost encryption strength, especially for color images, by ensuring diffusion occurs across color planes. Conceptually shown in \cref{3d_fission}, a pixel in one channel is influenced by neighboring pixels in other channels. For instance, during diffusion, a Green (G) channel pixel at $(i,j)$ is updated using nearby Red (R) channel pixels (\eg from $(i-1,j)$ and $(i,j-1)$), its previous G neighbor, and a key-derived element. A Blue (B) channel pixel may also rely on G channel values. This inter-channel dependency is vital to counteract independent channel analysis. TDA's operations and updates are detailed in \cref{Alg1}, using key-dependent matrices $X_{\text{2D}}$, $Y_{\text{2D}}$, and $Z_{\text{2D}}$.

\subsection{Decryption Algorithm}
\label{subsec:decryption}
The decryption procedure for TDADL-IE is designed as the exact inverse of the encryption process. Given that the core operations—chaotic sequence generation, index shuffling (permutation), and three-dimensional diffusion (typically involving operations such as XOR and modular arithmetic)—are deterministic and reversible when the correct secret keys ($Key_{x1}$, $Key_{y1}$, $Key_{a}$, $Key_{N_0}$) are known, the original plaintext image can be perfectly reconstructed. The decryption involves:
\begin{enumerate}
\item \textit{Chaotic Sequence Regeneration}: The same set of chaotic sequences (${s}$, $X_{\text{2D}}$, $Y_{\text{2D}}$, $Z_{\text{2D}}$) are regenerated using the secret keys identically to the encryption phase.
\item \textit{Inverse Three-dimensional Diffusion}: The TDA operations are applied in reverse. This typically involves processing in the reverse order of diffusion passes (\eg, inverse backward pass followed by inverse forward pass) and applying the mathematical inverse of each encryption operation (\eg, XOR is its own inverse; modular subtraction for modular addition). The cross-channel dependencies are also reversed according to the defined inverse rules.
\item \textit{Inverse Index Shuffling}: The inverse permutation, derived from the same chaotic sequence ${s}$ by determining the reverse mapping of indices, is applied to the diffused image to restore the original pixel order.
\end{enumerate}
The successful recovery of the plaintext image hinges on the precise application of these inverse steps and the availability of the identical secret keys used during encryption.

\section{Experimental results}
\label{Sec3}
This section evaluates the proposed cryptosystem, focusing on the efficiency and security of the TDADL-IE algorithm and the performance characteristics of the 1D-SQCM chaotic map.

\subsection{Experimental Setup}
The validation consists of analyzing the 1D Sin-Cos Pi Hyperchaotic Map (1D-SQCM) and evaluating the TDADL-IE encryption algorithm. The chaotic behavior of 1D-SQCM is compared to three one-dimensional chaotic maps \cref{t111} and two-dimensional systems. For TDADL-IE, three standard grayscale test images, "IMG1", "IMG2", and "IMG3" of sizes $M \times N$, \eg, and $512 \times 512$ pixels \cite{msoud_nickparvar_2021}, are used. It's compared with state-of-the-art encryption techniques \cite{luo2019image,zhang2019multiple,iqbal2023multi,lu2020efficient,kamal2021new} using standard metrics. Experiments ran on an AMD Ryzen 7 7800X3D CPU at 4.20 GHz, 32 GB RAM, with Windows 11 and MATLAB R2024a.

\subsection{Performance Analysis of the 1D-SQCM Chaotic Map}
This subsection presents a comparative analysis of the chaotic dynamics exhibited by the proposed 1D-SQCM and other existing chaotic maps, with detailed comparisons provided in \cref{t111}. The objective is to demonstrate the enhanced chaotic properties of 1D-SQCM.

\begin{table}[ht!]\footnotesize
	\centering
	\caption{Tested maps.}
	
	\begin{tabular}{llc}
		\toprule
		\textbf{Name} & \textbf{Mathematical expressions}   & \textbf{  Parameter settings }\\ \midrule
		
		1D-SQCM& $x_{n+1}=\sin(\frac{a}{\sin(x_{n})})$  & $a$$\in$(0, $\infty$)\\
		sine&$x_{n+1}=b\times \sin(\pi x_{n})  $&      $b$$\in$(0,  $\infty$)    \\	
		1-DFCS& $x_{n+1}=\frac{\cos((\eta x_{n}+1)^2+1)}{\sin((\eta x_n+1)^2+1)+2}$ & $\eta$$\in$(0, $\infty$) \\	
		1D-Chebyshev& 
		$x_{n+1}=\cos(k \times \arccos(x_{n}))$ & $k$$\in$(0, $\infty$)\\	
		\bottomrule
	\end{tabular}
	\label{t111}
\end{table}

\subsubsection{Lyapunov Exponent Analysis}
\begin{figure}[htb]
	\centering
	{
		\begin{minipage}[b]{1\linewidth}
			\centering
			\includegraphics[width=\linewidth]{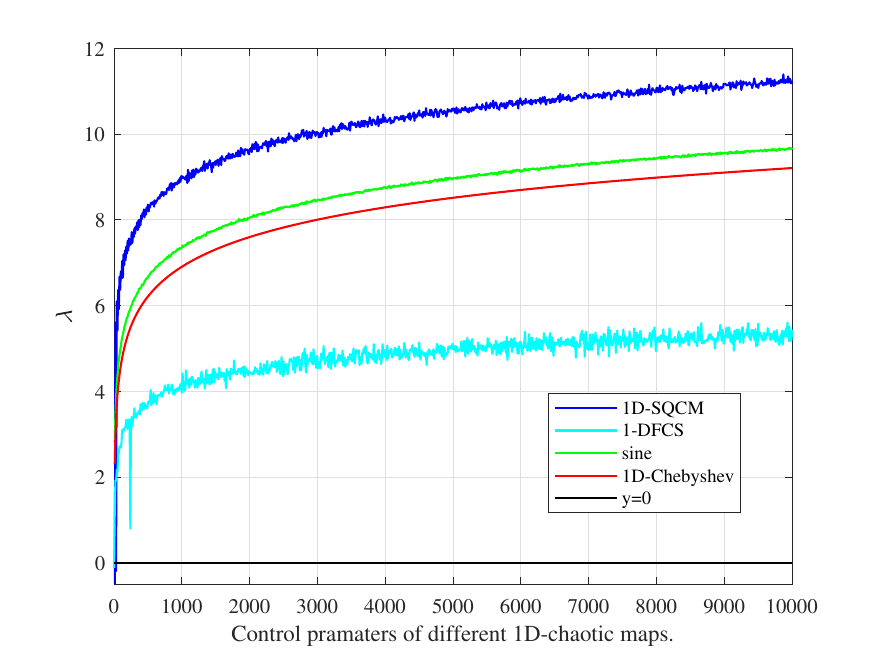}	
		\end{minipage}			
	}	
	
	\caption{The Lyapunov exponent of 1D-SQCM, 1-DFCS, Sine and 1D-Chebyshev map.}
	\label{1Dlyy}
\end{figure}

The Lyapunov exponent (LE) quantifies the sensitivity of a dynamical system to infinitesimal perturbations in its initial conditions, with positive LEs indicating chaotic behavior. \Cref{1Dlyy} illustrates the Lyapunov exponents for the 1D-SQCM and other benchmarked 1D chaotic maps across a control parameter range of $(0, 10000)$. The results demonstrate that the 1D-SQCM maintains positive Lyapunov exponents throughout this range. Furthermore, these values are consistently larger than those of the compared chaotic maps, signifying robust chaotic characteristics and heightened sensitivity to initial conditions, which are desirable for cryptographic applications.

\subsubsection{NIST SP 800-22 Statistical Test Suite}
The randomness of the sequences generated by the 1D-SQCM was rigorously evaluated using the NIST SP 800-22 statistical test suite. This suite comprises a set of tests designed to detect deviations from true randomness in binary sequences. \Cref{NIST} presents the P-values obtained from these tests applied to sequences derived from the 1D-SQCM output. For all generated sequences, the P-values consistently exceed the standard significance threshold of 0.01.

\begin{table}[ht!]\footnotesize
	\centering
	\caption{NIST test results of 1D-SQCM.}
	\begin{tabular}{llcc}
		\toprule
		\textbf{Test index} & \textbf{P$-$value } & \textbf{Result}    &            \\ \midrule
		\multirow{1}{*}{Frequency (Monobit) Test}    &   0.0909&PASS      &\\ 
		\multirow{1}{*}{Frequency Test within a Block}   &0.0711       &PASS        &  \\ 
		\multirow{1}{*}{Runs Test}       & 0.2368       &PASS       &  \\ 
		\multirow{1}{*}{Longest Run of Ones in a Block Test}       &0.4190     &PASS        &  \\ 
		\multirow{1}{*}{Binary Matrix Rank Test}    &0.7981   &PASS    &  \\ 
		\multirow{1}{*}{Discrete Fourier Transform (Spectral) Test}    &0.3504   &PASS    &  \\ 
		\multirow{1}{*}{Non-overlapping Template Matching Test}    &0.2368  &PASS    &  \\ 
		\multirow{1}{*}{Overlapping Template Matching Test}    &0.2757  &PASS    &  \\ 
		\multirow{1}{*}{Maurers Universal Statistical Test}    &0.9780  &PASS    &  \\ 
		\multirow{1}{*}{Linear Complexity Test}    &0.8676&PASS    &  \\ 
		\multirow{1}{*}{Serial Test}    &0.8343  &PASS    &  \\ 
		\multirow{1}{*}{Approximate Entropy Test}    &0.5341 &PASS    &  \\ 
		\multirow{1}{*}{Cumulative Sums (Cusums) Test}    &0.2896  &PASS    &  \\ 
		\multirow{1}{*}{Random Excursions Test}    &0.8623&PASS    &  \\ 
		\multirow{1}{*}{Random Excursions Variant Test}    &0.8676 &PASS    &  \\ 	
		\bottomrule
		
	\end{tabular}
	\label{NIST}
\end{table} 

\subsubsection{Deep Learning-based Predictability Assessment}

\begin{figure}[ht]
	\centering
	\includegraphics[width=\linewidth]{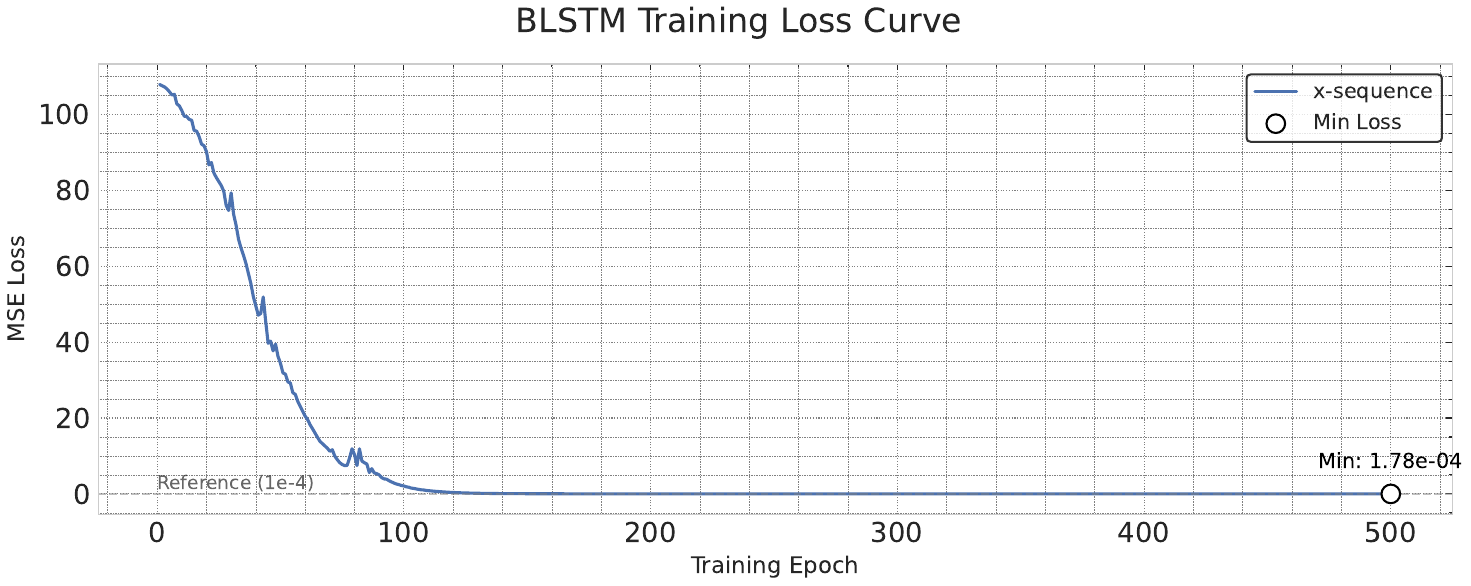}
	\caption{BLSTM training loss}
	\label{Gru training loss}
\end{figure}

% \begin{figure}[ht]
% 	\centering
% 	\includegraphics[width=\linewidth]{figs/combined_chaos_Mn_plot.pdf}
% 	\caption{The different 0-1 test K value of chaotic signals}
% 	\label{The difference of chaotic signals}
% \end{figure}

\begin{figure}[ht]
	\centering
	\includegraphics[width=\linewidth]{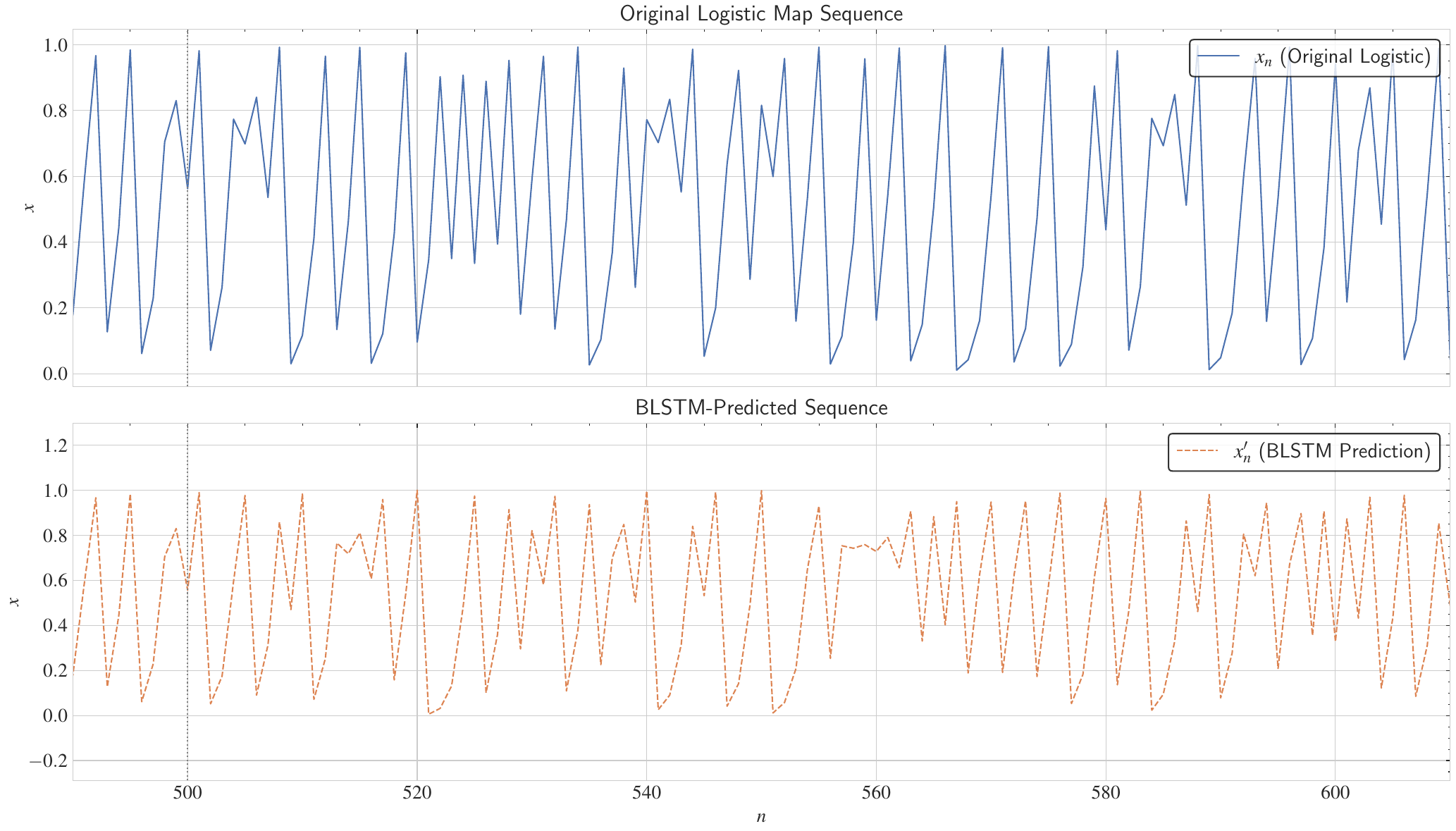}
	\caption{The difference of chaotic signals.}
    \label{diff}
	\label{Mn_true_vs_fully_predicted}
	%	\vspace{-10pt}
\end{figure}

The 1D-SQCM's output was analyzed using a Bidirectional Long Short-Term Memory (BLSTM) network, with training details in \cref{Mn_true_vs_fully_predicted}. The low root-mean-square error (RMSE) and loss function values suggest the BLSTM's ability to learn the short-term patterns of the chaotic sequence. A comparison between the actual sequence $\{X_{true}\}$ and the BLSTM prediction $\{X_{pred}\}$ is presented in \cref{diff}. The divergence between $\{X_{true}\}$ and $\{X_{pred}\}$ over extended predictions highlights the long-term unpredictability of the 1D-SQCM, a trait of robust chaotic systems.

\subsection{Security and Performance Analysis of TDADL-IE Cryptosystem}

\subsubsection{Key Space Analysis}
A fundamental requirement for robust cryptographic security against exhaustive search attacks is a key space substantially larger than $2^{100}$. The TDADL-IE cryptosystem utilizes four secret keys with the following initial theoretical ranges: $Key_{x1} \in (0, 1)$, $Key_{y1} \in (0, 1)$, $Key_{a} \in (0, \infty)$, and $Key_{N0} \in (2000, \infty)$. In a practical digital implementation, these continuous ranges are discretized based on the precision of the numerical representation. Consequently, the effective discrete key space is vast, estimated to be approximately $2^{208}$. This significantly surpasses the aforementioned security threshold, rendering brute-force attacks computationally infeasible and thereby providing a high level of protection. 

% \subsubsection{Histogram Analysis}
% To resist statistical attacks, an effective encryption algorithm must transform the original image's pixel distribution into one that lacks discernible patterns. \Cref{figure_histograms} presents the histograms of the original test images alongside their encrypted counterparts produced by TDADL-IE. The histograms of the encrypted images are observed to be remarkably uniform and substantially different from the typically non-uniform distributions of the original images. This flatness in the ciphertext histograms signifies that the encryption process evenly distributes pixel intensities, effectively masking the statistical characteristics of the plaintext and thereby bolstering security against statistical analysis attacks.

% \begin{figure}[h!]
% 	\centering
% 	\subfigure[]
% 	{\includegraphics[width=2.5cm,height=2.5cm]{Figure/1.png}}
%  \subfigure[]
%  {\includegraphics[width=2.5cm,height=2.5cm]{Figure/new/21.png}}
%  \subfigure[]
% 	{\includegraphics[width=2.5cm,height=2.5cm]{Figure/1.png}}
% 	\subfigure[]
%  {\includegraphics[width=2.5cm,height=2.5cm]{Figure/2.png}} 
% 	\subfigure[]
%  {\includegraphics[width=2.5cm,height=2.5cm]{Figure/new/22.png}}
% 	\subfigure[]
%  {\includegraphics[width=2.5cm,height=2.5cm]{Figure/2.png}} 
% 	\caption{Simulation results: (a,e) The original images of Image1, Image2, and Image3 respectively; (b,e) The encrypted images of (a,d), respectively; (c,f) The decrypted images of (b,f).}
% 	\label{figure_histograms}
% \end{figure}

\subsubsection{Differential Attack Analysis: Plaintext and Key Sensitivity}
An encryption algorithm's resilience to differential attacks is measured by its sensitivity to small changes in plaintext or key, assessed via NPCR and UACI metrics. Ideal NPCR and UACI values are about 99.6\% and 33.4\% for 8-bit grayscale images with random-like cipher images. The TDADL-IE algorithm's NPCR and UACI scores when altering a single bit are high and near these ideal levels, indicating strong diffusion properties and robustness to differential cryptanalysis \cref{compare}.
			
\begin{table}[ht]
    \centering
    \label{compare}
    \caption{Comparing coefficients of correlation schemes of one image.}
    \adjustbox{width=\linewidth}{
    \begin{tabular}{ccccccccc}
        \toprule
        \multirow{2}{*}{Algorithm} & \multicolumn{3}{c}{Correlation coefficients} & \multirow{2}{*}{Information entropy $\uparrow$} & \multirow{2}{*}{Time (s) $\downarrow$} & \multirow{2}{*}{NPCR} & \multirow{2}{*}{UACI} & \multirow{2}{*}{Key space $\uparrow$} \\ \cmidrule(lr){2-4}
        & H $\downarrow$ & V $\downarrow$ & D $\downarrow$ & & & & &  \\
        \midrule
        Proposed & \textbf{0.0011} & \textbf{-0.0009} & \textbf{0.0016} & \textbf{7.9993} & \textbf{0.3814} & \textbf{99.6094(*)} & \textbf{33.4641(*)} & \textbf{$2^{208}$} \\
	% \cite{lu2020efficient} &0.0024&	0.0011& 0.0021& 7.9971 & 1.4890& 99.6554& 33.4665& $2^{124}$\\
	Kamal \etal \cite{kamal2021new}&	0.0145&	0.0115&	0.0087&7.9992&3.0901&	99.6010&33.4389&$2^{115}$\\
    Ye \etal \cite{ye2022image}&	0.0085&	-0.0056&	0.0040&7.9992&3.3714&	99.6132&33.4268&$2^{156}$\\
     Xu \etal \cite{xu2019fast}&	0.0067&	-0.0086&	0.0140&7.9935&0.5509&	99.6102&33.4812&$2^{30}$\\

        \bottomrule
    \multicolumn{9}{l}{\footnotesize (*This value is closest to the ideal value)} \\
    \end{tabular}}
\end{table}

\subsubsection{Correlation Coefficient Analysis}
Plaintext images show high correlation among adjacent pixels. An effective encryption algorithm should reduce these correlations in ciphertext, preventing statistical attacks. \Cref{compare} shows that while original images have strong correlations, the TDADL-IE encrypted images have coefficients near zero in all directions. This reduction indicates that TDADL-IE effectively decorrelates pixel values, enhancing security against statistical cryptanalysis.

\subsubsection{Encryption Efficiency}
Computational efficiency is crucial for an image encryption algorithm's practicality. The TDADL-IE scheme's encryption speed was evaluated in the specified environment and compared with other algorithms in \cite{luo2019image,zhang2019multiple,iqbal2023multi,lu2020efficient,kamal2021new}. According to \Cref{compare}, it processes images competitively, proving its efficiency and suitability for applications requiring timely processing.

\section{Conclusion}
\label{Sec4}
This paper presents a novel color image encryption method addressing low key and plaintext sensitivity. The method uses a new one-dimensional chaotic system (1D-SQCM) and Two-Directional Diffusion Algorithm (TDA) for high security and efficiency. 1D-SQCM generates pseudo-random sequences, and the Index Shuffling Algorithm (ISA) rearranges pixel data in the permutation phase. The TDA ensures modifications in plaintext are diffused thoroughly, enhancing security and efficiency. Experiments show strong resilience against attacks, establishing the method as an effective and secure digital image encryption approach.

\bibliographystyle{IEEEtran} % 例如使用plainnat风格 
\bibliography{mycite}   % name your BibTeX data base

\end{document}